\documentclass[aps, pra, reprint, superscriptaddress, onecolumn, 11pt]{revtex4-2}

\usepackage{amssymb}
\usepackage{graphicx}
\usepackage{amsmath}
\usepackage{cancel}
\usepackage{bm}
\usepackage{tikz}
\usetikzlibrary{arrows.meta,positioning,calc,decorations.pathmorphing}
\usepackage{hyperref}
\usepackage{xcolor}
\usepackage[normalem]{ulem}

\hypersetup{colorlinks=true, linkcolor=blue, linktoc=page,
            citecolor=blue, urlcolor=blue}

\newcommand{\ze}[1]{{\color[rgb]{0.2,0,0.8}{#1}}}

\begin{document}

\title{Subsonic terahertz plasmon instability in space-time plasmonic crystals}

\author{H. Ter\c cas}
\email{hugo.tercas@isel.pt}
\affiliation{Instituto Superior de Engenharia de Lisboa, Instituto
Polit\'ecnico de Lisboa, 1959-007 Lisboa, Portugal}
\affiliation{GoLP/IPFN, Instituto Superior T\'ecnico, Universidade de Lisboa,
1049-001 Lisboa, Portugal}

\author{J. L. Figueiredo}
\email{jose.luis.figueiredo@tecnico.ulisboa.pt}
\affiliation{GoLP/IPFN, Instituto Superior T\'ecnico, Universidade de Lisboa,
1049-001 Lisboa, Portugal}

\author{P. Cosme}
\email{p.a.cosmeesilva@uva.nl}
\affiliation{Van der Waals-Zeeman Institute, Institute of Physics, University of Amsterdam, Amsterdam, The Netherlands}

\begin{abstract}

We investigate collective plasma-wave dynamics in a two-dimensional electron fluid subjected to a travelling periodic modulation of the effective plasma velocity. Using a hydrodynamic description combined with Floquet-Bloch theory, we show that the travelling modulation acts as a moving plasmonic crystal whose band structure is governed by an effective drift velocity in the comoving frame. We demonstrate that the closure of the Bragg gap at the edge of the Brillouin zone gives rise to an exceptional point and triggers a Krein-type instability through the coupling of positive- and negative-energy modes. Analytical expressions are derived for the instability threshold and growth rate, revealing that self-amplification can occur even when the electron drift remains subsonic in the laboratory frame. Numerical simulations of the nonlinear hydrodynamic equations confirm the theoretical predictions and show the evolution towards a saturated regime dominated by a coherent Floquet-Bloch mode. These results identify travelling plasmonic crystals as a novel platform for wave amplification and non-Hermitian band dynamics in electronic fluids.
\end{abstract}

\maketitle

\section{Introduction}
\label{sec:intro}

Manipulating waves through temporal variations of the medium has become a theme
across several branches of physics, for the reason that a temporal interface does
precisely the opposite of a spatial one: where the latter conserves frequency and
changes the wavevector, the former conserves the wavevector and changes the frequency, and
that single exchange of roles suffices to generate time refraction and time
reflection, frequency conversion, energy exchange with the medium, photonic time
crystals and Floquet-engineered propagation
\cite{MendoncaShukla2002,Mendonca2024,Galiffi2022,Engheta2023,BarHillel2024}. At the origin of these effects is the mixing of positive- and
negative-frequency components by a sudden change of a constitutive parameter,
from which a temporally transmitted and a temporally reflected wave emerge
\cite{MendoncaShukla2002}; interference between successive temporal events
follows naturally, and with it the notion of a temporal beam splitter
\cite{MendoncaMartins2003}, whose optical counterpart has since been observed in
the form of double-slit diffraction in time \cite{Tirole2023}. Modulating in
space and in time simultaneously promotes the medium to a spacetime
metamaterial \cite{CalozDeck2019,Mendonca2024}, and the program has been
extended to structured light, where time refraction acts upon orbital angular
momentum \cite{MendoncaOAM2026}, as well as to plasmonic systems, for which the
temporal response of the surface modes of a quantum plasma slab has been worked
out explicitly \cite{MendoncaFigueiredoTercas2026}.

Focusing on continuum theories, we can consider that the essential elements for spatiotemporal refraction are an oscillating field, thus described by a set of hyperbolic equations, and varying transport coefficients of the equations. As such, hydrodynamic theories (or quasi-conservation laws in a broader sense) are the natural language for such phenomena. In the present work, we are concerned with low-dimensional conductors well described by electron hydrodynamics where collective modes obey charged viscous fluid equations \cite{Lucas2018,Narozhny2019}, with the instability
predicted by Dyakonov and Shur \cite{Dyakonov1993,Dyakonov1996} playing an important role. Here, a current-carrying field-effect transistor with asymmetric boundary
conditions at the source and at the drain converts a stationary flow into a
coherent plasma oscillation, and has long been regarded as the most promising
route to a compact terahertz source \cite{Tonouchi2007,Dyakonov2010,Knap2008,Otsuji2013,Otsuji2014}. High-mobility two-dimensional materials
rendered the hydrodynamic regime experimentally accessible
\cite{Mendl2018,Crabb2021}, and in graphene devices this mechanism is known to
produce coherent frequency combs tunable through the gate voltage, the injected
current, the magnetic field and feedback
\cite{CosmeTercas2020,CosmeTercas2021,Cosme2024}. Additionally, replacing the single gate
by a periodic array promotes the channel to a plasmonic crystal with a
Bragg-folded band structure, \ze{and} the grating-gate structure constitutes a mature platform in which the gaps are tunable by bias
\cite{Koseki2016,Sai2023,Gorbenko2024} and in which the amplification of
terahertz radiation by a drifting electron fluid has been observed at room
temperature \cite{Boubanga2020}. The bulk instability of such a \emph{static}
crystal, however, requires a drift to surpass the plasma-wave velocity
\cite{Kachorovskii2012,Aizin2016,Petrov2017,Aizin2023}, which channels rarely
deliver, and it is arguably this circumstance that has confined the subject to
the cavity geometry of Dyakonov and Shur.

In this work, we show that a \emph{travelling} modulation lifts that restriction,
for the elementary reason that the quantity to which the fluid responds is the
drift measured in the frame of the modulation, so that a pattern running against
the current adds to it rather than competing with it. The travelling crystal is
then a static one at the effective drift $\widetilde u_0=u_0-\Omega/\kappa$; its Bragg gap
closes as soon as $|\widetilde u_0|$ exceeds the plasma-wave velocity. Also, the Bragg and Krein physics need references). Crucially, the modulation need not be imposed from outside. A second gated channel, serving as the gate of the first, carries a
plasmon whose own dispersion already supplies the required phase velocity, and a
Dyakonov-Shur cavity in that same channel converts a dc bias into the oscillation
that drives it, so that the entire chain is sustained by dc bias alone.

This paper is organized as follows. Section~\ref{sec:DS} shows how to treat the cavity not as a competing mechanism but as
the component that will supply the modulation, and establishes along the way why a \emph{uniform} temporal modulation, being blind to the drift, cannot help. Sec.~\ref{sec:crystal} develops the travelling-crystal formalism, from the Doppler
equivalence and the gap-to-instability transition to the threshold imposed by momentum relaxation and by viscosity, including the nonlinear state into which the instability saturates. Finally, in Sec.~\ref{sec:proto} we turn to more practical questions about set-up and implementation, including
which modulation depth a real channel is able to supply, which material survives that test, and how a single additional electron gas meets at once the two requirements of a travelling pattern at the plasmon velocity and of a source to sustain it.

\section{The Dyakonov-Shur cavity as an oscillator}
\label{sec:DS}

Consider a gated two-dimensional electron fluid of length $L$ in the hydrodynamic
regime, where the density $n(x,t)$ and the drift velocity $u(x,t)$ obey, in
the gradual-channel approximation \cite{Dyakonov1993,Dyakonov1996,Lucas2018,%
Narozhny2019},
\begin{align}
\partial_t n+\partial_x(nu)&=0,
\label{eq:cont} \\
\partial_t u+u\partial_xu&= -\frac{s_0^2}{n_0}\partial_x n -\frac{u}{\tau},
\label{eq:euler}
\end{align}
with $s_0=[e^2n_0/(mC)]^{1/2}$ the plasma-wave velocity, $C=\varepsilon_0 \varepsilon_r/d$ the gate capacitance per unit area, and $\tau$ the momentum relaxation time. Linearising
about a stationary current-carrying background $n=n_0$, $u=u_0$, itself sustained against the damping term $-u/\tau$ by a dc source, and neglecting dissipation for a moment, one obtains
\begin{equation}
(\partial_t+u_0\partial_x)^2\delta n = s_0^2\partial_x^2\delta n ,
\label{eq:waveeq}
\end{equation}
whose plane-wave solutions yield the dispersion relation
\begin{equation}
\omega_\pm=(u_0\pm s_0)k,
\label{eq:dispersion}
\end{equation}
for the downstream and the upstream plasmons, respectively. The momentum relaxation contributes with the additional term $\tau^{-1}(\partial_t+u_0\partial_x)\delta n$ on the left of Eq.~(\ref{eq:waveeq}) and, for $\omega\tau\gg1$, damps both branches with   rate $1/(2\tau)$. What renders the configuration unstable is the asymmetry of the boundary conditions imposed at the source and at the drain \cite{Dyakonov1993,%
Dyakonov1996},
\begin{equation}
\delta n(0,t)=0, \qquad
\delta(nu)|_{x=L}\equiv n_0\delta u|_{x=L}+u_0\delta n|_{x=L}=0,
\label{eq:bcs}
\end{equation}
which couple the two branches of Eq.~(\ref{eq:dispersion}). Writing
$\delta n(x)=A_+e^{ik_+x}+A_-e^{ik_-x}$, with $k_\pm=\omega/(u_0\pm s_0)$ and using
$\delta u_\pm=\pm(s_0/n_0)\delta n_\pm$, Eqs.~(\ref{eq:bcs}) require the conditions
$A_++A_-=0$ and $(s_0+u_0)A_+e^{ik_+L}=(s_0-u_0)A_-e^{ik_-L}$, such that
\begin{equation}
e^{i(k_+-k_-)L} = -\frac{s_0-u_0}{s_0+u_0},
\label{eq:DScondition}
\end{equation}
whose real part contains the modes $\text{Re} \, 
\omega_\ell = (2\ell+1)\pi/\tau_\text{cyc}$. For $0<u_0<s_0$, the wave is amplified on every round trip. Defining the round-trip gain $g=(s_0+u_0)/(s_0-u_0)$ and the round-trip time $\tau_{\rm cyc}=2Ls_0/(s_0^2-u_0^2)$, the
amplitude grows as $g^N$ after $N$ cycles and the identification $g=e^{\gamma_{\rm DS}\tau _{\rm cyc}}$ yields the Dyakonov-Shur growth rate,
\begin{equation}
\gamma_{\rm DS} = \frac{s_0^2-u_0^2}{2s_0L}
 \ln\!\left( \frac{s_0+u_0}{s_0-u_0} \right) ,
\label{eq:gammaDS}
\end{equation}
reaching a maximum of $0.448\,s_0/L$ at $u_0=0.648\,s$ and  being active whenever $\gamma_{\rm DS}>1/2\tau$. Therefore, the quasi-modes form an equidistant comb, $\omega_\ell=[(2\ell+1)\pi]/\tau_{\rm cyc} +i\gamma_{\rm DS}$. Two features of Eq.~(\ref{eq:gammaDS}) will be exploited in Sec.~\ref{sec:proto}: the instability requires no external drive, and a dc
bias to hold $u_0$ against friction on a channel short enough that $0.448\,s_0/L>1/2\tau$ suffices to produce a coherent oscillation at $\omega_0=\pi/\tau_{\rm cyc}$, which is an on-chip terahertz oscillator driven by dc bias alone. The frequency is fixed by the cavity length, which can be regarded as an electronic Fabry-P\'erot cavity.

Before turning to a travelling modulation it is worth consider modulating the plasma-wave velocity with (Floquet) period $T_F$, such that $s_0$ is replaced by the time-dependent plasmon velocity $s(t)$ satisfying the property $s(t+T_F)=s(t)$, as in time-refraction phenomena \cite{MendoncaShukla2002, Mendonca2024, BarHillel2024}. 
%
For $u_0=0$, eliminating $\delta u$ gives
\begin{equation}
  \partial_t^2\delta n(x,t)+\tau^{-1}\partial_t\delta n(x,t)=s^2(t)\,\partial_x^2\delta n(x,t)  , \label{2_deltan}
\end{equation}
together with the boundary conditions of Eqs.~(\ref{eq:bcs}). The spatial eigenfunctions $\sin k_\ell x$, with $k_\ell =(\ell +\tfrac12)\pi/L$, are therefore independent of time and thus, according to Eq.~\eqref{2_deltan}, the coefficients of $\delta n=\sum_\ell q_\ell(t)\sin k_\ell x$ obey a Hill equation of the form
\begin{equation}
    \ddot q_\ell +\frac{\dot q_\ell}{\tau }+s^2(t)k_\ell ^2q_\ell =0.
\end{equation}
Here, the friction is removed upon the standard substitution $q_\ell\to e^{-t/2\tau}q_\ell$, which leaves an undamped Hill equation with $s^2k_\ell^2\to s^2k_\ell^2-1/4\tau^2$. 

Consider, for definiteness, a square wave alternating between $s_1$ and $s_2$ with period $T_F$, such that $s=s_1$ for $0<t<T_F/2$ and $s=s_2$ for $T_F/2<t<T_F$. Within each half-period, $s$ is constant and the undamped Hill equation is solved by $q_\ell (t)=q_\ell(0)\cos\omega_i t+\dot q_\ell(0)\,\omega_i^{-1}\sin\omega_it$, so that, defining $X=(q_\ell,\dot q_\ell)^{T}$, we may write the unitary evolution as
\begin{equation}
X(t)=U_{i,\ell}(t)\,X(0),\qquad
U_{i,\ell}(t)=
\begin{pmatrix}
\cos\omega_{i,\ell}t & \omega_{i,\ell}^{-1}\sin\omega_{i,\ell}t \\[2pt]
-\omega_{i,\ell}\sin\omega_it & \cos\omega_{i,\ell}t
\end{pmatrix},
\qquad
\omega_{i,\ell}=\left(s_i^2k_\ell ^2-\frac{1}{4\tau^2}\right)^{1/2},
\label{eq:propagator}
\end{equation}
with $i=1,2$. The square modulation therefore consists in the application of two successive unitaries,  $X(T_F)=U_{2,\ell}(T_F/2)\,U_{1,\ell}(T_F/2)\,X(0)\equiv\mathcal M_\ell X(0)$, and the monodromy matrix for each mode $q_\ell $ is
\begin{equation}
\mathcal M_\ell=
\begin{pmatrix}
\cos\theta_{2, \ell} & \omega_{2,\ell}^{-1}\sin\theta_2 \\[2pt]
-\omega_{2,\ell}\sin\theta_{2,\ell} & \cos\theta_{2,\ell}
\end{pmatrix}
\begin{pmatrix}
\cos\theta_{1,\ell} & \omega_{1,\ell}^{-1}\sin\theta_{1,\ell} \\[2pt]
-\omega_{1,\ell}\sin\theta_{1,\ell} & \cos\theta_{1,\ell}
\end{pmatrix}, 
\qquad
\theta_{i,\ell}=\frac{\omega_{i,\ell} T_F}{2}.
\label{eq:monodromy}
\end{equation}
Since $\det\mathcal M_\ell =1$, the mode is unstable whenever
$|{\rm Tr}\,\mathcal M_\ell |>2$, and the associated growth rate is $\gamma_\ell=\mu_\ell -1/2\tau$, where $\mu_\ell =T_F^{-1}\,{\rm Re}\left[{\rm arcosh}({\rm Tr}\,\mathcal M_\ell /2)\right]$ denotes the Floquet exponent \cite{NayfehMook1979, Tercas_2010}. In explicit terms, equation~(\ref{eq:monodromy}) gives,
\begin{equation}
{\rm Tr}\,\mathcal M_\ell/2=\cos(\theta_{1,\ell}+\theta_{2,\ell})
-2|\beta_\ell|^2\sin\theta_{1,\ell}\sin\theta_{2,\ell},
\end{equation} 
with $|\beta_\ell|^2=(\omega_1-\omega_2)^2/4\omega_1\omega_2$. The quantity $\beta_\ell$ is the time-reflection coefficient of a sudden switch of the mode frequency from $\omega_1$ to $\omega_2$ \cite{MendoncaShukla2002}: each of the two temporal interfaces per period converts a fraction $\beta_\ell$ of the
positive-frequency wave into a negative-frequency one. For $\omega_{i,\ell}\tau\gg1$, one may set $\omega_{i,\ell} \simeq s_ik_\ell$, whereupon $|\beta_\ell|^2\simeq|\beta|^2 =(s_1-s_2)^2/4s_1s_2$ becomes independent of the mode. The instability
condition $|{\rm Tr}\,\mathcal M_\ell|>2$ is met in tongues around $\theta_{1,\ell}+\theta_{2,\ell}=p\pi$, $p=1,2,\dots$, the widest of which, the
principal parametric resonance $p=1$, reads $\bar sk_\ell=\pi/T_F$ with
$\bar s=(s_1+s_2)/2$. In these terms, to first order in $|s_1-s_2|$, we obtain the maximum parametric gain of the $\ell$-th mode as
\begin{equation}
\gamma_\ell ^{\max}\simeq\frac{|s_1-s_2|\,k_\ell}{\pi}-\frac{1}{2\tau}.
\label{eq:Hillmax}
\end{equation}
This means that, by time modulation, the channel acquires an additional amplification channel. However, a difficulty remains since this particular form of modulation has no short-wavelength cutoff: noticing that $|{\rm Tr}\,\mathcal M_\ell/2|\le1+2|\beta|^2$, the Floquet exponent is bounded by $\mu_\ell T_F\le2\,{\rm arsinh}|\beta|$, independently of $k_\ell $, so that short-wavelength modes are not less unstable than the cavity modes. A cutoff must be supplied by the viscosity of the electron fluid, and will be introduced in Sect.~\ref{sec:threshold}. Finally, the principal resonance requires driving the whole channel at twice the cavity frequency, which itself lies in the terahertz source, making it unpractical. Below, we try to circumvent these difficulties by constructing a space-time plasmonic crystal. 

\section{Space-time plasmonic crystal}
\label{sec:crystal}

Let the plasma-wave velocity be in the form of a travelling wave by modulating the channel capacity $C(x,t)$,
\begin{equation}
s^2(x,t)=s_0^2\left[1+\epsilon\cos(\kappa x-\Omega t)\right],
\label{eq:s2xt}
\end{equation}
with the quantity $V_{0} =\Omega /\kappa $ defining the modulation velocity and $\epsilon$ the modulation amplitude. The electrostatic field perturbation becomes $\delta\phi=-e\,\delta n/C(x,t)$, such that the linearised dynamics assumes the form
\begin{align}
\partial_t \delta n+u_0\partial_x \delta n+n_0\partial_x \delta u&=0,\nonumber\\
\partial_t \delta u+u_0\partial_x \delta  u+\partial_x(s \,\delta n)&=-\frac{\delta u}{\tau }.
\label{eq:ab}
\end{align}
In the Lagrange frame, $\xi=x-V_0t$, the modulation is static and Eqs.~(\ref{eq:ab}) keep
their form under the transformations
\begin{equation}
\tilde  u_0=u_0-V_0,\qquad \tilde \omega =\omega+kV_0 .
\label{eq:u0p}
\end{equation}
A travelling modulation is nothing other than a static crystal seen at the effective drift $\tilde u_0$, with the Floquet replicas $(\omega+\ell \Omega,k+\ell\kappa)$, with $\ell$ an integer,  being Doppler copies of one and the same static band structure. The change of frame also removes the time ordering discussed above. Indeed, in the comoving frame, the linear operator does not depend on time, so the Floquet problem reduces to an ordinary Bloch eigenvalue problem and no monodromy matrix is needed. Below, we investigate the practical consequences of the space-time modulation.

We now introduce the characteristic variables $w_\pm=s\delta n\pm n_0\delta u$, in terms of which Eqs.~(\ref{eq:ab}) read 
\begin{equation}
\left[\frac{\partial}{\partial t}+(\tilde u_0\pm s)\frac{\partial}{\partial \xi}\right]w_\pm
=-\frac{\epsilon\kappa s_0^2}{4s}\sin\kappa\xi\left(\frac{\tilde u_0}{s}\mp1\right)(w_++w_-).
\label{eq:char}
\end{equation}
Expanding $(\delta n, \delta u)$ as $\sum_m(a_m,b_m)\,e^{i(k+m\kappa)\xi-i\omega t}$ and the modulation as $s^2(\xi)=\sum_pS_pe^{ip\kappa\xi}$, with
\begin{equation}
S_p=\frac{\kappa}{2\pi}\int_0^{2\pi/\kappa}s^2(\xi)\,e^{-ip\kappa\xi}\,d\xi
=s_0^2\left[\delta_{p,0}+\frac{\epsilon}{2}\left(\delta_{p,1}+\delta_{p,-1}\right)\right],
\label{eq:Sp}
\end{equation}
and eqs.~(\ref{eq:ab}) produce $\omega a_m=k_m\widetilde u_0 a_m+k_mb_m$ and $\omega b_m=k_m\widetilde u_0 b_m+k_m\sum_lS_{m-l}a_l$, with $k_m=k+m\kappa$. After eliminating $b_m$, we get
\begin{equation}
(\omega-k_m\tilde u_0)^2a_m=k_m^2\sum_\ell S_{m-\ell}\,a_\ell =k_m^2s_0^2\left[a_m+\frac{\epsilon}{2}\left(a_{m-1}+a_{m+1}\right)\right].
\label{eq:harmonics}
\end{equation}
Because $S_p$ vanishes for $|p|>1$, each harmonic is coupled only to its nearest neighbours, $m\pm1$. Keeping the lowest modes only, $m=0,-1$, yields
\begin{equation}
\left[\widetilde \omega-\omega_+(k)\right]\left[\widetilde \omega-\omega_-(k-\kappa)\right]=-\frac{(\epsilon s_0)^2}{16}\,k(k-\kappa).
\label{eq:twowave}
\end{equation}
The branches cross at $k_c=(s_0-\widetilde u_0)\kappa /2s_0$, where
\begin{equation}
\omega=\omega_c  \pm\frac{\epsilon \kappa}{8}\sqrt{s_0^2-\widetilde u_0^2},\qquad
\omega_c=\frac{(s_0^2-\widetilde u_0^2)\kappa}{2s_0}.
\label{eq:gap}
\end{equation}
For $|\widetilde u_0|<s_0$, this gives rise to a (Bragg) gap and the frequencies are real; the gap closes at $|\widetilde u_0|=s_0$, signaling an exceptional point \cite{Heiss2012, MiriAlu2019}, and for $|\widetilde u_0|>s_0$ the square root becomes imaginary, and the corresponding growth rate $\gamma =\rm Im(\omega)$ is
\begin{equation}
\gamma=\frac{\epsilon \kappa}{8}\sqrt{\widetilde u_0^2-s_0^2},\qquad |\widetilde u_0|>s_0 ,
\label{eq:gamma_tw}
\end{equation}
which is different from the Dyakonov-Shur growth rate $\gamma_{\rm DS}$ in Eq. \eqref{eq:gammaDS}. The mechanism at work here is a {\it Krein collision} \cite{Cairns1979, KollarMiller2014}: for $\widetilde u_0>s_0$, $k_c(k_c-\kappa)>0$ in Eq.~(\ref{eq:twowave}) and the two coupled branches carry wave energy of opposite sign, a circumstance for
which a sufficient condition of stability follows from
\begin{equation}
E=\frac{1}{2}\int\left[n_0\,\delta u^2+\frac{s^2}{n_0}\,\delta n^2
+2\widetilde u_0\,\delta n\,\delta u\right]d\xi ,
\label{eq:energy}
\end{equation}
which is conserved in the crystal frame for $\tau\to\infty$ and positive definite
for $|\widetilde u_0|<s_{\min}=s_0\sqrt{1-\epsilon}$. All modes then have positive
energy and the spectrum is real. Negative-energy (unstable) modes, instead, require
$|\widetilde u_0|>s_{\min}$.

The crossing $k_c$ moves with the drift, and reaches the centre of a Brillouin
zone, $k_c\equiv0 \pmod\kappa$, whenever
\begin{equation}
\widetilde u_0=(2p+1)\,s_0,\qquad p=0,1,2,\dots
\label{eq:zonecentre}
\end{equation}
The case $p=0$ is the exceptional point itself, where $\gamma=0$, so that the first
useful zone centre is $\widetilde u_0=3s_0$, with $k_c=-\kappa$. Only there does the
unstable Bloch mode carry a uniform, $k=0$, component and radiate: a gated plasmon
has a phase velocity of order $s_0$, two to three orders of magnitude below $c$, so
that every other component lies outside the light cone.

\subsection{Effect of viscosity and plasmonic crystal band structure}
\label{sec:threshold}

As stated, momentum relaxation shifts the dispersions by $-i/2\tau$ and Eq.~(\ref{eq:gamma_tw}) would give the instability threshold
$|u_0-\Omega/\kappa|>s_0[1+(4/\epsilon s_0\kappa\tau)^2]^{1/2}$. A more complete treatment of the problem requires including a kinematic viscosity $\nu\partial_x^2u$ to Eq.~(\ref{eq:euler}). In that case, the Krein collision couples $k_c$ to its Bragg partner at $k_c-\kappa$, which damps at a rate $\nu\kappa^2$. Viscosity thus modifies each branch with some additional damping, $\Gamma_i=(1/\tau+\nu k_i^2)/2$, with $k_1=k_c=(s_0-\tilde u_0)\kappa/2s_0$ and $k_2=k_c-\kappa =-(s_0+\tilde u_0)\kappa /2s_0$. Since the two coupled branches carry wave energy of opposite sign, the determinant of the coupled problem at the crossing is $\gamma^2-\Gamma_1\Gamma_2$ ($\gamma$ being the lossless rate of Eq.~(\ref{eq:gamma_tw})), and a Krein collision grows when the coupling exceeds the geometric mean of the two dampings, the net rate being $-\bar\Gamma+(\gamma_0^2+\Delta\Gamma^2/4)^{1/2}$ with $\bar\Gamma=(\Gamma_1+\Gamma_2)/2$ and $\Delta\Gamma=\Gamma_1-\Gamma_2$. Explicitly, the instability is therefore active provided the condition
\begin{equation}
\frac{\epsilon \kappa}{8}\sqrt{\tilde u_0^2-s_0^2}
\;>\;\frac12\left[\left(\frac{1}{\tau}+\frac{\nu\kappa^2}{4}\frac{(\tilde u_0-s_0)^2}{s_0^2}\right)
\left(\frac{1}{\tau}+\frac{\nu\kappa^2}{4}\frac{(\tilde u_0+s_0)^2}{s_0^2}\right)\right]^{1/2} .
\label{eq:threshold}
\end{equation}
The gain is linear in the modulation wavenumber $\kappa$, while the viscous loss is quadratic; viscosity bounds the crystal from the short-period side as momentum relaxation bounds it from the long-period side, and thus an optimum period exists.

\begin{figure}[t!]
\includegraphics[width=\columnwidth]{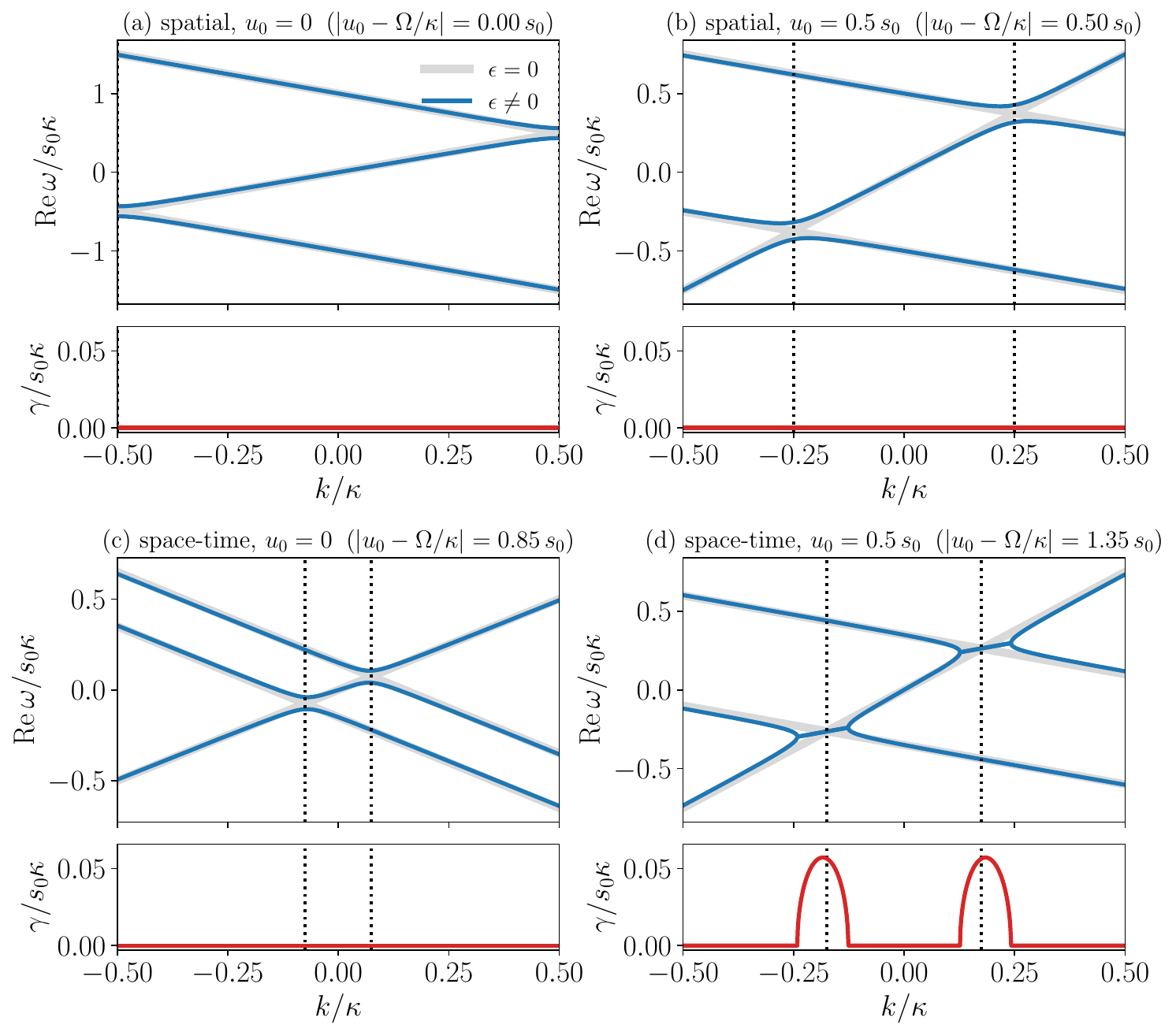}
\caption{Band structure of the plasmonic crystal in the reduced zone,
$|k|<\kappa/2$, for $\epsilon=0.5$, from the three-band model of
Eq.~(\ref{eq:threeband}), in the laboratory frame, $\widetilde\omega=\omega+kV_0$, and
in units of $s_0\kappa$. Upper plots, ${\rm Re}\,\widetilde\omega$ of the dressed
modes (blue) with the unmodulated bands (grey); lower plots, the growth rate
$\gamma=\max\,{\rm Im}\,\widetilde\omega$, which is the same in both frames. Dotted
lines mark the Bragg crossings $\pm k_c$. (a) Static crystal, $V_0=0$, without
drift, $\widetilde u_0=0$. (b) Static crystal with $u_0=0.5\,s_0$: the drift tilts
the bands and moves the crossings, but the gaps stay open and $\gamma=0$.
(c) Space-time crystal with $V_0=-0.85\,s_0$ and $u_0=0$, so that
$\widetilde u_0=0.85\,s_0<s_0$ and the gaps are still open. (d) The same crystal at
the drift of (b): $\widetilde u_0=1.35\,s_0>s_0$, the gaps have closed at
exceptional points and $\gamma\neq0$ in a band around $\pm k_c$, with
$\gamma_{\max}\simeq0.057\,s_0\kappa$ within the three-band model. At this depth
that value exceeds by $47\%$ the converged plane-wave result,
$\gamma_{\max}=0.039\,s_0\kappa$, which moreover peaks at $|k|\simeq0.27\,\kappa$,
outside the arcs.}
\label{fig:bands}
\end{figure}

Keeping the three harmonics $m=0,\pm1$, the downstream plasmon $\omega_+(k)$ couples
to the two upstream plasmons folded by $\mp\kappa$, $\omega_-(k\mp\kappa)$, and the
dressed bands obey
\begin{multline}
\left[\omega-\omega_+(k)\right]\left[\omega-\omega_-(k-\kappa)\right]\left[\omega-\omega_-(k+\kappa)\right]\\
=-\frac{(\epsilon s_0)^2}{16}\left\{k(k-\kappa)\left[\omega-\omega_-(k+\kappa)\right]+k(k+\kappa)\left[\omega-\omega_-(k-\kappa)\right]\right\},
\label{eq:threeband}
\end{multline}
with $\omega_\pm(k)=(\widetilde u_0\pm s_0)k$. The upstream branch folded by
$-\kappa$ crosses $\omega_+$ at $k=k_c$ and the one folded by $+\kappa$ at $k=-k_c$, so
that Eq.~(\ref{eq:threeband}) treats both crossings of the reduced zone at once;
discarding the harmonic $m=+1$ removes the second and recovers
Eq.~(\ref{eq:twowave}). Figure~\ref{fig:bands} shows the resulting bands in the
laboratory frame, $\widetilde\omega=\omega+kV_0$, for $\epsilon=0.5$ and four
combinations of drift and modulation velocity, chosen so that $\widetilde u_0$
increases from panel to panel: a static crystal ($V_0=0$) without drift,
$\widetilde u_0=0$, and with $u_0=0.5\,s_0$, $\widetilde u_0=0.5\,s_0$
[Figs.~\ref{fig:bands}(a) and (b)]; and a crystal travelling against the
current at $V_0=-0.85\,s_0$, without drift, $\widetilde u_0=0.85\,s_0$, and with the
same drift as in (b), $\widetilde u_0=1.35\,s_0$ [Figs.~\ref{fig:bands}(c) and
(d)]. In the first three $|\widetilde u_0|<s_0$, the gaps are open and
$\mathrm{Im}\,\omega=0$: neither the drift alone nor the modulation alone
destabilises the crystal. In the fourth, drift and modulation velocity add to
$\widetilde u_0>s_0$, the gaps close at the exceptional points, the real parts of
the two coupled branches merge, and an arc of positive $\mathrm{Im}\,\omega$
opens around each crossing, as Eq.~(\ref{eq:gamma_tw}) requires. The two arcs are
mapped onto each other by the symmetry $(k,\omega)\to(-k,-\omega^*)$ of real
fields.

\subsection{Nonlinear saturation of the instability}
\label{sec:nonlinear}

Equations~(\ref{eq:gap})--(\ref{eq:threshold}) are first order in $\epsilon$ and
linear in the perturbation. It is, however, important to understand how accurate the two-wave result is at usable modulation depths, and what the instability saturates into. To get some numerical insight, we integrate the full nonlinear fluid model within the gradual-channel approximation,
\begin{align}
\partial_t n+\partial_x(nu)&=0,
\label{eq:nlcont}\\
\partial_t u+u\,\partial_xu&=-\frac{1}{n_0}\,\partial_x\!\left[s^2(x,t)\,(n-n_0)\right]
-\frac{u-u_0}{\tau}+\nu\,\partial_x^2u ,
\label{eq:nleuler}
\end{align}
with $s^2(x,t)$ from Eq.~(\ref{eq:s2xt}). We integrate pseudo-spectrally in $x$ with $2/3$ dealiasing and fourth-order Runge-Kutta in time \cite{Orszag1971,Canuto2006}, in a periodic box of eight crystal periods with $64$ points per period and $\delta t=0.01\,(s_0\kappa)^{-1}$ up to $t=600\,(s_0\kappa)^{-1}$, from Gaussian white noise of rms amplitude $10^{-6}n_0$ in the density and $10^{-6}s_0$ in the velocity. We run directly at the operating point of the device: $\epsilon=0.5$, $V_0=-2.846\,s_0$ and $u_0=0.154\,s_0$, hence $\widetilde u_0=3.00\,s_0$, with $\tau=97\,(s_0\kappa)^{-1}$ and $\nu=0.056\,s_0\kappa^{-1}$.

\begin{figure}[t!]
\centering
\includegraphics[width=\linewidth]{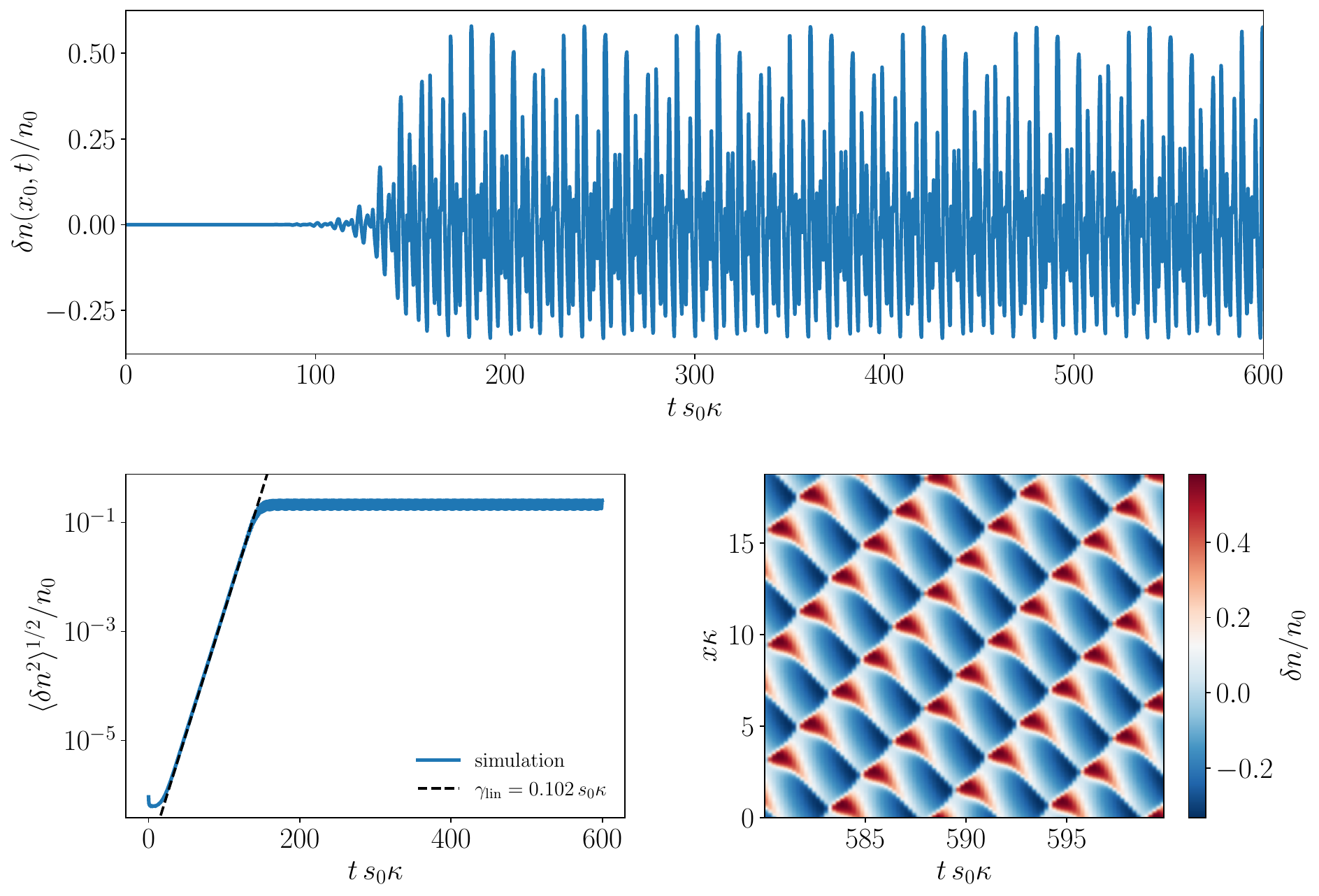}
\caption{Nonlinear evolution, from Eqs.~(\ref{eq:nlcont})--(\ref{eq:nleuler}), at the operating point of Sec.~\ref{sec:device}: $\epsilon=0.5$, $V_0=-2.846\,s_0$, $u_0=0.154\,s_0$, hence $\widetilde u_0=3.00\,s_0$, with $\tau=97\,(s_0\kappa)^{-1}$ and $\nu=0.056\,s_0\kappa^{-1}$, in a box of eight periods seeded with white noise of rms amplitude $10^{-6}n_0$ in the density and $10^{-6}s_0$ in the velocity. (a) Density perturbation at the centre of the box. (b) Root-mean-square perturbation (solid) and the exponential of the linear growth rate, $\gamma=0.102\,s_0\kappa$ (dashed), of the unstable Bloch wave; the growth departs from the exponential near $0.1\,n_0$ and saturates at $0.222\,n_0$. (c) Space-time map of $\delta n/n_0$ over the last $20\,(s_0\kappa)^{-1}$ and three crystal periods: the Bloch wave of Eq.~(\ref{eq:blochwave}) is locked to the modulation and travels upstream at $V_0$, and its steep fronts are the higher Bragg harmonics.}
\label{fig:nlcrystal}
\end{figure}

The evolution of the electron density fluctuations is shown in Fig.~\ref{fig:nlcrystal}. The noise organises itself into the zone-centre Bloch mode, the one at which the two branches of opposite energy are in resonance, and grows exponentially at the rate of the linear problem, with viscosity dominating the loss. The growth departs from the exponential once the rms density modulation reaches $\langle \delta n^2\rangle^{1/2} \sim0.1\,n_0$, as advection steepens the wave into the sharp fronts of Fig.~\ref{fig:nlcrystal}(c), which feed the short wavelengths where the viscosity acts. This nonlinear transfer balances the gain at $\langle\delta n^2\rangle^{1/2}\simeq0.22\,n_0$. Deep into the nonlinear stage, the instability is arrested at finite amplitude, and what remains is a coherent wave locked to the modulation.

\begin{figure}[t!]
\centering
\includegraphics[width=\linewidth]{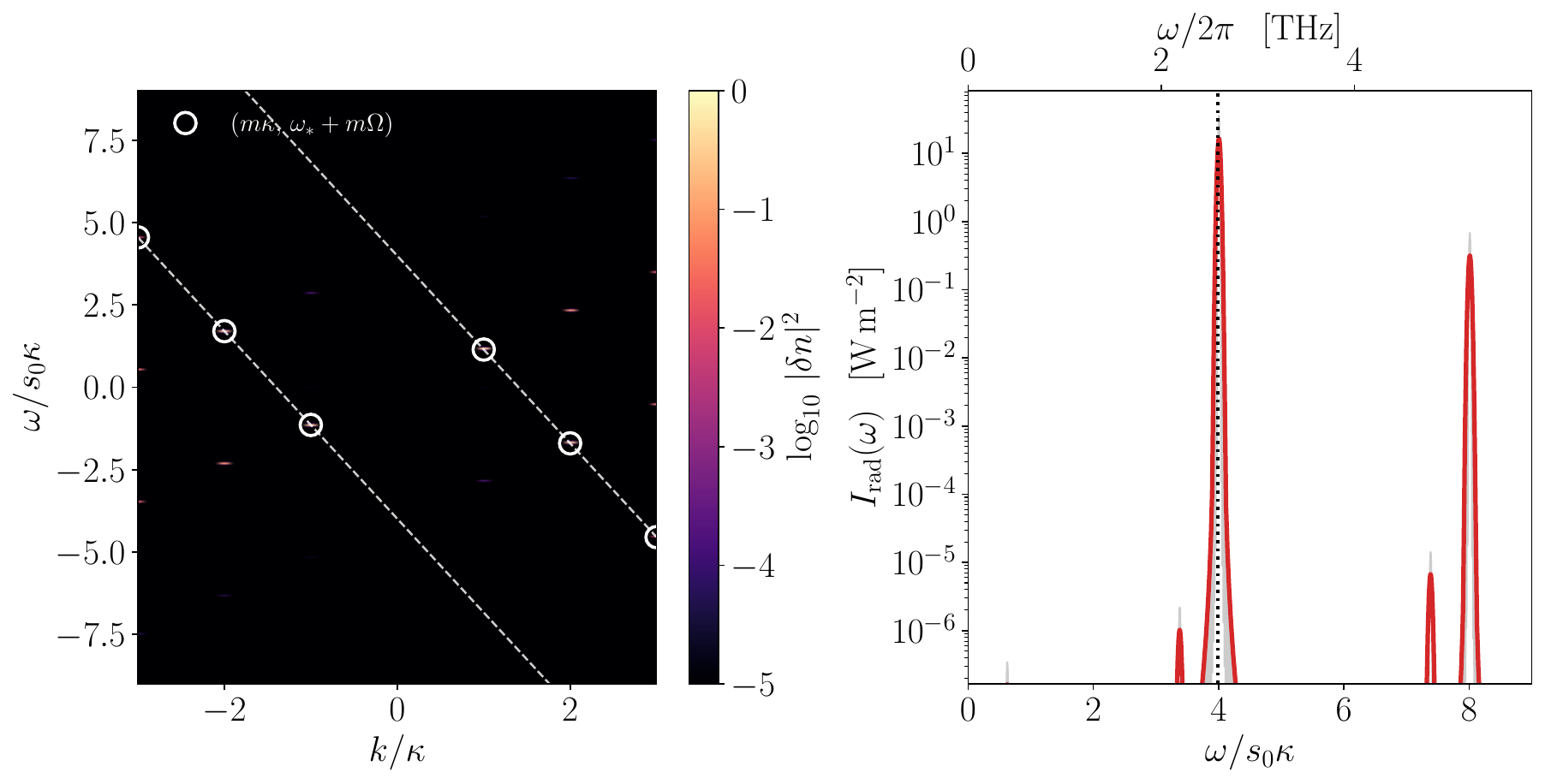}
\caption{Saturated state of the run of Fig.~\ref{fig:nlcrystal}, over the window
$t>240\,(s_0\kappa)^{-1}$, with a Hann window in time and fourfold zero padding,
in the laboratory frame. (a) Power $|\delta n(k,\widetilde\omega)|^2$ normalised to
its maximum. The dashed lines are $\widetilde\omega=\pm\omega_*+kV_0$, with
$\omega_*=3.990\,s_0\kappa$ the crystal-frame frequency of the linearly unstable
zone-centre mode, and the circles mark their populated members $m=-1,-2,-3$;
$m=0$ is empty because mass conservation forbids a uniform density component, and
the $m>0$ members are empty because the Bragg coupling at $\widetilde u_0>s_0$ is
one-sided. Both lines have the same slope $V_0=-2.85\,s_0$, which is negative because the
modulation runs against the current. (b) Radiated intensity of the uniform sheet current
$j=-e\langle nu\rangle_x$, raw (grey) and Gaussian filtered (red), for an
unscreened sheet between vacuum and the GaAs substrate; the upper axis is the
frequency in THz. The dotted vertical is $\widetilde\omega$. The comb is harmonic
in $\widetilde\omega$ and contains nothing at $\Omega$ or at
$\widetilde\omega\pm\Omega$, those lying below $10^{-12}$ of the fundamental.}
\label{fig:nlspectrum}
\end{figure}

The spectrum of Fig.~\ref{fig:nlspectrum} shows the same in Fourier space. The power does not spread over the $(k,\widetilde\omega)$ plane: it collects on two parallel straight lines of slope $V_0$, that is, at a single frequency in the frame of the modulation. The saturated state is therefore one nonlinear Floquet-Bloch wave,
\begin{equation}
\delta n(x,t)={\rm Re}\left[e^{-i\widetilde\omega t}\sum_m a_m\,e^{im(\kappa x-\Omega t)}\right],
\qquad \widetilde\omega=4.006\,s_0\kappa ,
\label{eq:blochwave}
\end{equation}
oscillating at the centre of the Brillouin zone ($k=0$), with the second and third harmonics added by the nonlinearity at $2\widetilde\omega$ and $3\widetilde\omega$. The density has zero dc component, but the current does, and it is the uniform current that radiates, since only waves with $k\approx0$ lie inside the light cone. The current $j(t)=-e\langle nu\rangle_x$, with $\langle\cdot\rangle_x$ the spatial average, is quadratic in the fields and comes from the beating of the harmonics of Eq.~(\ref{eq:blochwave}): its component at $\widetilde\omega$ arises from the beating of the fundamental with its second harmonic, that is, at third order in the amplitude. Because the Bloch wavenumber is zero there is no Doppler shift, and the emission is a comb at $\widetilde\omega_\ell =\ell \widetilde\omega$ containing both even and odd harmonics, as expected in a system without inversion symmetry, the dc drift and the travelling modulation each singling out a direction. Nothing is emitted at $\Omega$ or at $\widetilde\omega\pm\Omega$, because those components of the wave have wavenumbers $m\kappa$, outside the light cone: for a prescribed and infinitely extended modulation the pump is invisible at the output (the finite device is discussed in Sec.~\ref{sec:limits}).

A uniform sheet current of amplitude $j_0$, radiating into vacuum on one side and into a substrate of refractive index $n_r=\sqrt{\varepsilon_r}$ on the other, emits the intensity $I_{\rm rad}=Z_0\,j_0^2/[2(1+n_r)]$, where $Z_0\simeq377\,\Omega$ is the impedance of vacuum. For GaAs ($n_r\simeq3.6$) the fundamental gives $36$~W\,m$^{-2}$. This is the strength of the source, not the power of a finished emitter. The metal gate, at $h=109$~nm from the channel (Sec.~\ref{sec:device}), acts as a ground plane and suppresses the intensity by $(2n_rk_0h)^2=1.8\times10^{-3}$, with $k_0=\widetilde\omega/c$, unless it is opened over the emitting region; and the sheet formula assumes a crystal wider than the free-space wavelength, $\lambda_0=116\,\mu$m, that is, at least $116$ periods for $a=1\,\mu$m. Both belong to the out-coupling calculation, which we do not attempt (Sec.~\ref{sec:limits}).

\section{Performance estimates based on a self-pumped prototype}
\label{sec:proto}

Everything so far has taken Eq.~(\ref{eq:s2xt}) as a given minimal model for the parametric modulation of the sound speed. In a real device, however, the performance depends on the ability of the medium to produce autonomously two things at once: a pattern of the requisite depth, and a pattern travelling at a velocity of order $s_0$. In the gradual-channel approximation the sound speed of the plasmonic perturbations combines two contributions, one from the Fermi pressure and another from the Coulomb interaction governed by the gate capacitance. Yet a gate cannot change its own capacitance dynamically at the frequencies required; what a gate voltage alters is the local carrier density, from which it follows that Eq.~(\ref{eq:s2xt}) is an idealisation of the real modulation---a point to which we return in Sec.~\ref{sec:limits}.

As regards the depth, it is enough to write the plasma-wave velocity of a gated fluid in full,
\begin{equation}
s_0^2=\frac{v_{\rm F}^2}{2}+\frac{e^2n_0d}{m\varepsilon_0\varepsilon_r},
\label{eq:s0full}
\end{equation}
the compressibility of the degenerate fluid plus the gate term already introduced in Sec.~\ref{sec:DS}, which is the only one a gate voltage can reach. For $n=n_0[1+\delta\cos(\kappa x-\Omega t)]$, with $\delta\le1$ since the density cannot go negative, the two kinds of band behave differently. In a parabolic band $m$ is fixed and $v_{\rm F}^2=2\pi\hbar^2n/m^2$, so that both terms of Eq.~(\ref{eq:s0full}) are proportional to $n$ and $s_0^2$ follows the density exactly, the first harmonic being exactly $\epsilon=\delta$. In a Dirac band it is $v_{\rm F}$ that is fixed, with $m=\hbar\sqrt{\pi n}/v_{\rm F}$, so that the compressibility term does not respond at all while the gate term grows only as $\sqrt n$, and a full swing of the density produces only half a swing of $s_0^2$. To first order in $\delta$,
\begin{equation}
\epsilon=
\begin{cases} \left(1-\dfrac{v_{\rm F}^2}{2s_0^2}\right)\dfrac{\delta}{2}, & \text{Dirac band},\\[10pt] \delta, & \text{parabolic band}.\end{cases}
\label{eq:depth}
\end{equation}
Far from being a detail, the factor of one half is decisive. In graphene at $d=5$~nm the gate term carries $79\%$ of $s_0^2$, so that Eq.~(\ref{eq:depth}) caps $\epsilon$ at $0.40$, and no gate thickness lifts the cap above $0.50$; the cap is itself a first-order estimate, the exact first harmonic of $(1+\delta\cos)^{1/2}$ being $0.600$ rather than $0.500$ at $\delta=1$, but a modulation that deep empties the channel at the troughs, and $\delta\le0.9$, for which $\epsilon\le0.40$, is the most that can reasonably be assumed. Against that, the critical depth of Eq.~(\ref{eq:threshold}), minimised over the period, is $0.80$ at room temperature and $0.56$ at $150$~K, at both zone centres of Eq.~(\ref{eq:zonecentre}). Graphene is therefore short by a factor of two in the first case and by some forty per cent in the second, and nothing in the modulation technique recovers the difference---the very dispersion that gives graphene its long relaxation time and its room-temperature hydrodynamics is what halves the available depth. The comparison is collected in Table~\ref{tab:materials}: the room-temperature III-V channels possess the depth but not the time, $\tau\simeq0.25$--$0.26$~ps leaving them stable at every period and at any admissible depth, and silicon, with $\tau\simeq0.04$~ps, fares worse still, so that a single entry survives, and it is cryogenic.
\begin{table}[t!]
\begin{ruledtabular}
\begin{tabular}{lccccc}
channel & band & $s_0$ ($10^6$m/s) & $\tau$ (ps) & $\epsilon_{\max}$ & $\epsilon_c$ \\
\colrule
Graphene/hBN, 300 K \cite{Mayorov2011,Bandurin2016}  & Dirac     & 1.56 & 1.17 & 0.40 & 0.80 \\
Graphene/hBN, 150 K \cite{Wang2013,KrishnaKumar2017} & Dirac     & 1.56 & 2.33 & 0.40 & 0.56 \\
AlInAs/InGaAs, 300 K \cite{Knap2004}                 & parabolic & 1.54 & 0.26 & 1.00 & --   \\
AlGaN/GaN, 300 K \cite{Ambacher2000,ElFatimy2010}    & parabolic & 1.48 & 0.25 & 1.00 & --   \\
Si MOSFET, 300 K \cite{KnapSi2004,Tauk2006}          & parabolic & 0.47 & 0.04 & 1.00 & --   \\
AlGaAs/GaAs, 40 K \cite{Umansky2009,Gusev2018}       & parabolic & 0.65 & 23.8 & 1.00 & 0.18 \\
\end{tabular}
\end{ruledtabular}
\caption{Supply and demand on the modulation depth. $\epsilon_{\max}$ is the largest depth a density modulation can reach, Eq.~(\ref{eq:depth}) at $\delta=1$, which for a parabolic band saturates the formal bound and should be read as an upper limit rather than an operating point; $\epsilon_c$ is the smallest depth that satisfies Eq.~(\ref{eq:threshold}) at $\widetilde u_0=3\,s_0$, minimised over the period for the first five entries, with a dash where no admissible depth works at any period. Graphene is quoted at $d=5$~nm with $\nu=0.10$~m$^2$/s, and GaAs at the $d=10$~nm of Sec.~\ref{sec:device}; for GaAs the minimum, $0.10$, occurs at $a\simeq3.3\,\mu$m, and the entry quoted is the value at the design period $a=1\,\mu$m. The relaxation times are those reported for the channels cited in the first column.}
\label{tab:materials}
\end{table}

\subsection{The two-channel device}
\label{sec:device}

\begin{figure}[t!]
\centering
\resizebox{\linewidth}{!}{%
  \definecolor{cA}{RGB}{31,119,180}
  \definecolor{cB}{RGB}{214,39,40}
  \definecolor{cG}{RGB}{90,90,90}
  \begin{tikzpicture}[>={Stealth[length=2.6mm]},line width=0.7pt]

  \begin{scope}
  \def\xL{2.2}\def\xR{12.0}\def\aa{3.267}     

  \def\ysub{1.00}\def\ysig{1.28}\def\ybar{1.98}\def\ypmp{2.44}\def\ygat{3.50}\def\ytop{3.90}

  \fill[cG] (\xL,\ygat) rectangle (\xR,\ytop);
  \node[font=\footnotesize,anchor=west] at (\xR+0.35,{(\ygat+\ytop)/2}) {metal gate};

  \fill[black!4] (\xL,\ypmp) rectangle (\xR,\ygat);

  \foreach \i in {0,...,195}{
    \pgfmathsetmacro{\xa}{\xL+\i*0.05}
    \pgfmathsetmacro{\fr}{50+50*cos(360*(\xa-\xL)/\aa)}
    \fill[cB!\fr!cA] (\xa,\ypmp) rectangle (\xa+0.051,{\ypmp-(\ypmp-\ybar)});
  }
  \draw[black!70,line width=0.9pt] (\xL,\ybar) rectangle (\xR,\ypmp);
  \foreach \j in {0,1,2}{
    \node[font=\scriptsize,white] at ({\xL+0.05+\j*\aa},{(\ybar+\ypmp)/2}) {$+$};
    \node[font=\scriptsize,white] at ({\xL+0.05+\aa/2+\j*\aa},{(\ybar+\ypmp)/2}) {$-$};
  }
  \node[font=\footnotesize,anchor=west,align=left] at (\xR+0.35,{(\ybar+\ypmp)/2})
     {pump strip,\ $s_p$};

  \foreach \i in {0,...,195}{
    \pgfmathsetmacro{\xa}{\xL+\i*0.05}
    \pgfmathsetmacro{\fr}{8+60*(0.5+0.5*cos(360*(\xa-\xL)/\aa))}
    \fill[cA!\fr!white] (\xa,\ysig) rectangle (\xa+0.051,\ybar);
  }
  \draw[black!35] (\xL,\ysig) rectangle (\xR,\ybar);

  \fill[black!30] (\xL,\ysub) rectangle (\xR,\ysig);
  \draw[black!70,line width=0.9pt] (\xL,\ysub) rectangle (\xR,\ysig);
  \draw[->,line width=1.1pt] (6.0,{(\ysub+\ysig)/2}) -- (7.6,{(\ysub+\ysig)/2});
  \node[font=\scriptsize,fill=white,inner sep=1.2pt,anchor=west] at (7.7,{(\ysub+\ysig)/2}) {$u_0$};
  \node[font=\footnotesize,anchor=west,align=left] at (\xR+0.35,{(\ysub+\ysig)/2})
     {signal channel,\ $s_0$};

  \fill[black!8]  (\xL,-0.45) rectangle (\xR,\ysub);
  \draw[black!35] (\xL,-0.45) rectangle (\xR,\ysub);
  \node[font=\footnotesize,black!60] at (8.6,0.25) {substrate};

  \def\xd{1.15}
  \foreach \yy in {\ysig,\ybar,\ypmp,\ygat}{\draw[dotted,black!45] (\xd,\yy) -- (\xL,\yy);}
  \draw[<->] (\xd,\ypmp) -- (\xd,\ygat)
     node[midway,anchor=east,font=\scriptsize,inner sep=2pt] {$d_p$};
  \draw[<->] (\xd,\ysig) -- (\xd,\ybar)
     node[midway,anchor=east,font=\scriptsize,inner sep=2pt] {$d$};

  \foreach \xc in {\xL-0.22,\xR-0.18}{
    \fill[black!65] (\xc,{\ysub-0.12}) rectangle (\xc+0.40,{\ysig+0.12});
    \fill[black!65] (\xc,{\ybar-0.12}) rectangle (\xc+0.40,{\ypmp+0.12});
  }
  \node[font=\scriptsize,anchor=south] at ({\xL-0.02},{\ypmp+0.18}) {dc};
  \node[font=\scriptsize,anchor=north] at ({\xL-0.02},{\ysub-0.18}) {dc};

  \draw[->,line width=1.2pt,cB] (7.3,2.95) -- (5.3,2.95);
  \node[font=\footnotesize,anchor=west,cB] at (7.45,2.95)
     {$V_0=-s_p$,\ \ $\Omega=-s_p\kappa$};
  \draw[<->] (\xL+0.05,2.52) -- ++(\aa,0)
     node[midway,font=\scriptsize,fill=black!4,inner sep=1pt] {$a$};

  \draw[cB,line width=1.1pt,decorate,
        decoration={snake,amplitude=1.1pt,segment length=5pt,post length=3.5pt},->]
     (4.6,1.02) -- (3.4,-0.85);
  \node[font=\footnotesize,cB,anchor=north] at (3.4,-0.90) {$\widetilde\omega$};

  \node[font=\bfseries] at (0.1,{\ytop+0.3}) {(a)};
  \end{scope}

  \begin{scope}[yshift=-3.1cm,xshift=1.6cm]
  \def\sc{3.3}\def\y{0}

  \fill[cA!12] (0,\y-0.30) rectangle ({1.0*\sc},\y+0.30);
  \fill[cB!14] ({1.0*\sc},\y-0.30) rectangle ({3.45*\sc},\y+0.30);
  \node[font=\scriptsize,cA!55!black] at ({0.5*\sc},\y) {gap open};
  \node[font=\scriptsize,cB!60!black] at ({2.2*\sc},\y) {Krein instability};

  \draw[->] (0,\y-0.30) -- ({3.62*\sc},\y-0.30)
     node[anchor=west,font=\footnotesize] {$\widetilde u_0/s_0$};
  \foreach \v in {0,1,2,3}{
    \draw ({\v*\sc},\y-0.30) -- ({\v*\sc},\y-0.42)
      node[anchor=north,font=\scriptsize,inner sep=2pt] {$\v$};}
  \draw[dashed,black!70] ({1.0*\sc},\y+0.30) -- ({1.0*\sc},\y+0.60);
  \node[font=\footnotesize,anchor=south,inner sep=1pt] at ({1.0*\sc},\y+0.56) {$s_0$};

  \draw[->,line width=1.1pt] (0,\y+1.25) -- ({0.154*\sc},\y+1.25);
  \draw[->,line width=1.1pt,cB] ({0.154*\sc},\y+1.25) -- ({3.0*\sc},\y+1.25);
  \draw[black!60] ({0.154*\sc},\y+1.12) -- ({0.154*\sc},\y+1.38);
  \node[font=\scriptsize,anchor=south,inner sep=1pt] at ({0.08*\sc},\y+1.33)
    {$u_0$};
  \node[font=\scriptsize,anchor=south,cB,inner sep=2pt] at ({1.6*\sc},\y+1.30)
    {$|V_0|=s_p$};
  \draw[dashed,cB] ({3.0*\sc},\y+0.30) -- ({3.0*\sc},\y+1.72);
  \node[font=\scriptsize,anchor=south,cB,align=center,inner sep=1pt] at ({3.0*\sc},\y+1.66)
    {$k_c=-\kappa$,\ zone centre};

  \node[font=\bfseries] at (-1.6,\y+1.85) {(b)};
  \end{scope}

  \end{tikzpicture}%
}
\caption{The self-pumped space-time plasmonic crystal, not to scale, with the
parameters only; their values are given in Sec.~\ref{sec:device}. (a) Two
tunnel-isolated two-dimensional electron gases in one GaAs/AlGaAs
heterostructure. The lower one is the signal channel, gated by the upper one
across $d$, which through Eq.~(\ref{eq:s0full}) fixes its plasma-wave velocity
$s_0$, and carrying the dc drift $u_0$. The upper one is the pump strip, gated
by the metal electrode across $d_p>d$, hence with the faster velocity $s_p>s_0$.
Each channel carries a dc bias and nothing else is applied. The Dyakonov-Shur
instability of Sec.~\ref{sec:DS}, operating in the pump strip, turns its dc
current into a coherent plasmon of wavevector $\kappa=2\pi/a$; the charge pattern
of that plasmon ($\pm$) travels at $V_0=-s_p$, so $\Omega=-s_p\kappa$ is fixed by
the strip's own dispersion rather than imposed, and its potential modulates the
plasma-wave velocity of the signal channel through the thin barrier (shading).
(b) What the signal channel responds to is $\widetilde u_0=u_0-V_0$, so the dc
drift and the pump velocity add head to tail. Tuned to the first useful zone
centre of Eq.~(\ref{eq:zonecentre}) the Bragg crossing sits at $k_c=-\kappa$,
where the unstable Bloch mode acquires the $k=0$ component that lets it radiate
at $\widetilde\omega$ through the substrate.}
\label{fig:setup}
\end{figure}

The device is two tunnel-isolated two-dimensional electron gases in one GaAs/AlGaAs heterostructure, both at $n_0=9.1\times10^{11}$~cm$^{-2}$, differing only in how far their gate sits (see Fig.~\ref{fig:setup} for a schematic representation). The lower gas is the signal channel and the upper gas is its gate, $d=10$~nm away, which by Eq.~(\ref{eq:s0full}) gives $s_0=0.648\times10^6$~m/s. The upper gas is the pump strip and a metal electrode is its gate, $d_p=99$~nm away, giving $s_p=1.845\times10^6$~m/s$\,=2.85\,s_0$, each carrying a dc bias and nothing else being applied to the device whatsoever.

By construction, the pump strip disposes of the harder of the two problems. A plasmon in it with wavevector $\kappa$ has frequency $|\Omega|=s_p\kappa$ not as an imposed matching condition but as its own dispersion relation, so that its potential pattern travels at $V_0=-s_p$ with neither delay line nor phase shifters, to the approximation, discussed in Sec.~\ref{sec:limits}, of a pump strip of negligible drift and linear dispersion; and this, rather than any question of convenience, is why the modulation is entrusted to a second electron gas instead of a metallic array, since a coplanar line on GaAs has a phase velocity near $c/3$ and bringing it down to $s_p\simeq c/160$ would require slowing it by a factor of about fifty, which is the real obstruction behind the apparently simpler phased-gate route. A gated electron gas is already that slow. The drift budget then closes on the zone centre: with $u_0$ at the saturation velocity, $10^5$~m/s, one has $u_0/s_0=0.154$ and $\widetilde u_0=u_0+s_p=3.00\,s_0$, which satisfies Eq.~(\ref{eq:zonecentre}) with $p=1$, so that $k_c=-\kappa$ and the unstable mode radiates.

It is viscosity that sets the operating temperature. For a GaAs electron gas the shear-stress relaxation time obeys $1/\tau_2=A_{ee}T^2[\ln(E_{\rm F}/T)]^2+1/\tau_{2,0}$ with $\tau_{2,0}=0.8\times10^{-11}$~s and $A_{ee}=0.9\times10^9$~s$^{-1}$K$^{-2}$, and $\nu=v_{\rm F}^2\tau_2/4$, while momentum relaxation obeys $1/\tau=A_{\rm ph}T+1/\tau_0$ with $A_{\rm ph}=10^9$~s$^{-1}$K$^{-1}$ and $\tau_0=5\times10^{-10}$~s, all measured on a gas of this density over $1.5$--$40$~K \cite{Gusev2018}. Viscosity falls with temperature because $\tau_2$ shortens, and the hydrodynamic condition improves for precisely the same reason, so that the two requirements align and the design calls for the warm end of the fitted range. At $T=40$~K this gives $\nu=5.8\times10^{-3}$~m$^2$/s, $\tau=23.8$~ps and an electron-electron mean free path of $56$~nm.

With $a=1\,\mu$m, the Bragg frequency is $\omega_{\rm B}/2\pi=s_0/a=0.65$~THz and every constraint is met with margin. The signal channel is hydrodynamic, $\kappa\ell_{ee}=0.35$, that is $a/\ell_{ee}=18$; it is also deep in the gated regime, $\kappa d=0.06$, whereas for the pump strip $\kappa d_p=0.62$ and the dispersion is only approximately the linear one the argument needs, as discussed in Sec.~\ref{sec:limits}; and Eq.~(\ref{eq:threshold}) asks for $\epsilon_c=0.18$ against the $\epsilon_{\max}=1$ that Eq.~(\ref{eq:depth}) supplies. Taking $\epsilon=0.5$, comfortably inside both, the plane-wave problem gives $\gamma/\omega_{\rm B}=0.1024$, that is a net growth rate of $4.2\times10^{11}$~s$^{-1}$ and a gain time of $2.4$~ps, with the maximum at the zone centre $k=0$ as designed. The value is converged in the Bloch truncation at $|m|\le4$ and unchanged to five digits at $|m|\le12$, $20$ and $28$, so the harmonics that set it reach only $k\ell_{ee}\simeq1$---at the edge of the hydrodynamic window rather than outside it, which is the price of a modulation this deep. The laboratory frequency of that mode is $3.99\,\omega_{\rm B}$, that is $2.59$~THz, and it leaves through the substrate.

The pump strip must of course itself oscillate, and it does. A Dyakonov-Shur cavity of length $L_p$ has its fundamental at $k=\pi/2L_p$, so that demanding $k=\kappa$ fixes $L_p=a/4=250$~nm, and then $|\Omega|=s_p\kappa$ follows with no tuning: $|\Omega|/2\pi=1.85$~THz. At that length Eq.~(\ref{eq:gammaDS}) gives $\gamma_{\rm DS}=4.0\times10^{11}$~s$^{-1}$. Its loss is dominated not by momentum relaxation, $1/2\tau=2.1\times10^{10}$~s$^{-1}$, but by viscosity, which at $k=\kappa$ damps at $\nu\kappa^2/2=1.1\times10^{11}$~s$^{-1}$: the margin against the sum of the two is three, not the nineteen that friction alone would suggest. The chain is in this way closed, from dc bias to Dyakonov-Shur oscillation at $1.85$~THz in the pump strip, thence to a travelling modulation of the signal channel, thence to a Krein instability at the zone centre, and thence to radiation at $2.59$~THz.

\subsection{Experimental limitations}
\label{sec:limits}

Several matters are left open, of which the geometry of the pump is the most
consequential. A Dyakonov-Shur cavity of length $L_p$ has the eigenmodes
$k_\ell=(2\ell+1)\pi/2L_p$ of Sec.~\ref{sec:DS}, so that the choice $L_p=a/4$
places the fundamental at $k_0=\kappa$ and the first overtone at $3\kappa$, and
the selection of the former is in fact secured by the same viscosity that limits
the crystal. Since $\gamma_{\rm DS}$ is common to the whole comb whereas the loss
$1/2\tau+\nu k^2/2$ grows quadratically, the fundamental survives with
$4.0\times10^{11}$ against $1.4\times10^{11}$~s$^{-1}$ while the overtone is
damped (a loss rate of $1.0\times10^{12}$~s$^{-1}$ exceeds the gain by a factor of two and a
half). What is not secured is the length. The round-trip gain of
Eq.~(\ref{eq:gammaDS}) falls as $1/L_p$ while the loss does not depend on the
cavity at all, so that oscillation requires
\begin{equation}
L_p<\frac{s_p^2-u_0^2}{2s_p}\ln\!\left(\frac{s_p+u_0}{s_p-u_0}\right)
\left(\frac{1}{2\tau}+\frac{\nu\kappa^2}{2}\right)^{-1}=0.74\,\mu\text{m},
\label{eq:Lmax}
\end{equation}
whereas the sheet formula of Sec.~\ref{sec:nonlinear} presumes a crystal wider
than the free-space wavelength, $\lambda_0=116\,\mu$m, that is $116$ periods. The
two requirements are a factor of some one hundred and sixty apart, and a single
cavity cannot meet both: the pump would have to be an array of about $460$
cavities of $250$~nm, mutually phase-locked to within a fraction of
$\Omega^{-1}$, and the mechanism of that locking is not addressed here.

The mode of each cavity is, moreover, a standing wave, $A_+=-A_-$, and a standing
modulation is the superposition of two travelling ones of half the depth,
\begin{equation}
s^2=s_0^2\left[1+\epsilon\cos\kappa x\cos\Omega t\right]
=s_0^2\left[1+\frac{\epsilon}{2}\cos(\kappa x-\Omega t)
+\frac{\epsilon}{2}\cos(\kappa x+\Omega t)\right],
\label{eq:standing}
\end{equation}
so that the crystal is driven at $\epsilon/2$ and not at $\epsilon$. The penalty
is severe but not fatal. At $\epsilon=0.5$ the counter-propagating half drives the
zone-centre mode at $\gamma=0.022\,s_0\kappa$ instead of the $0.102\,s_0\kappa$ of
a purely travelling modulation of the same depth, a reduction by a factor of
four and a half that leaves the available $\epsilon/2=0.25$ against the
$\epsilon_c=0.18$ of Table~\ref{tab:materials}, a margin of only $1.4$. The
co-propagating half, for its part, presents the signal channel with
$\widetilde u_0=u_0-s_p=-2.69\,s_0$, which exceeds $s_0$ in modulus and would be
unstable in the lossless two-wave estimate, but which at this depth lies below the
threshold of Eq.~(\ref{eq:threshold}), $\gamma=-0.010\,s_0\kappa$, and in any case
sits at $k_c=1.85\,\kappa$, away from a zone centre and therefore unable to
radiate. Each cavity is, finally, itself a Dyakonov-Shur emitter carrying a net
oscillating current at the pump frequency, so that the absence of $\Omega$ from
the output established in Sec.~\ref{sec:nonlinear}, which rests on the pump
residing entirely at wavenumbers $m\kappa$ outside the light cone, holds for a
prescribed and infinitely extended modulation and not necessarily for the device.

The zone-centre condition of Eq.~(\ref{eq:zonecentre}) is, next, a tight one.
Radiation requires the Bloch wavenumber to lie inside the light cone, and since
$k_c=-\kappa(\widetilde u_0/s_0-1)/2$ a departure $\delta\widetilde u_0$ from the
design value displaces it by $\delta k_c=-\kappa\,\delta\widetilde u_0/2s_0$, so
that the admissible detuning is
\begin{equation}
\frac{|\delta\widetilde u_0|}{s_0}\lesssim\frac{2\sqrt{\varepsilon_r}\,
\widetilde\omega}{c\,\kappa}\simeq0.06
\label{eq:tolerance}
\end{equation}
for emission into the substrate, and $0.017$ for emission into vacuum. The budget
$\widetilde u_0=u_0+s_p=3.00\,s_0$ neglects two shifts of that order or larger.
The pump strip carries its own dc drift, $u_{0p}\simeq10^5$~m/s for the same bias,
so that the phase velocity of its upstream plasmon is $u_{0p}-s_p$ rather than
$-s_p$ and $\widetilde u_0$ falls to $2.85\,s_0$; and with $\kappa d_p=0.62$ the
gated dispersion is no longer linear, the gate term of Eq.~(\ref{eq:s0full})
acquiring a factor $\tanh(\kappa d_p)/\kappa d_p=0.89$ while the compressibility
term is untouched, which lowers $s_p$ to $1.74\times10^6$~m/s. With both
corrections $\widetilde u_0=2.69\,s_0$ and $k_c=-0.84\,\kappa$, far outside
Eq.~(\ref{eq:tolerance}). The condition can be restored by retuning the gate,
$d_p\simeq136$~nm accommodating both ($\kappa d_p=0.85$, where the dispersion is
less linear still), but it must then be solved together with the two-channel
electrostatics, and the tolerance it imposes on the fabrication is of the order of
a few per cent.

\section{Conclusions}
\label{sec:concl}

In this work, we have shown that a travelling modulation of the plasma-wave
velocity provides a route to plasmonic instabilities in two-dimensional
electron fluids that does not require a supersonic current. In the frame moving
with the modulation, the system is a static plasmonic crystal in which the
electrons drift at $\widetilde u_0 = u_0 - \Omega/\kappa$, so that a modulation
propagating against the current adds to the dc drift. The Bragg gap of this
crystal closes at $|\widetilde u_0| = s_0$, and for larger drifts the two coupled
plasmon branches, which carry wave energy of opposite sign, become unstable with
the growth rate of Eq.~(\ref{eq:gamma_tw}). Nonlinear simulations show that the
instability saturates at an rms density modulation of $22\%$, in the form of a
single Floquet--Bloch wave locked to the modulation and travelling with it, whose
nonlinear harmonics carry the uniform current that radiates.

Carrying that to a device brings into play two quantities that the linear theory
does not contain, and one opportunity that it does not exploit. Viscosity bounds
the crystal from the short-period side as momentum relaxation bounds it from the
long-period side, Eq.~(\ref{eq:threshold}), leaving a finite window rather than a
free choice of period. The depth that a density modulation can actually reach,
Eq.~(\ref{eq:depth}), carries a factor of one half in a Dirac band, which rules
graphene out at room temperature, where the available $0.40$ falls short of the
required $0.80$ by a factor of two, in spite of its being the superior channel by
every other measure; the margin narrows to some forty per cent by $150$~K, so
that moderate cooling, rather than any change of modulation technique, is what
would bring a Dirac channel within reach. And the modulation itself need not be
supplied from outside, since a second gated electron gas, serving as the gate of
the first, possesses exactly the slow dispersion that the phase matching
requires, while a Dyakonov-Shur cavity within it converts a dc bias into the
oscillation that drives it.

What results is a two-channel GaAs/AlGaAs device at $40$~K of $1\,\mu$m period,
biased by dc alone, in which a $1.85$~THz pump generated on chip drives a
zone-centre Krein instability that grows in $2.4$~ps under an ideal travelling
modulation, or in $11$~ps once the standing character of a cavity mode is
accounted for, and that saturates into a coherent wave radiating a harmonic comb
at $2.59$~THz and its multiples, the pump frequency itself being absent from the
output of the idealised infinite crystal since the Floquet sidebands lie outside
the light cone. These statements rest on different footings and the distinction
is worth preserving: that a travelling modulation activates a Bloch instability
through a collision of opposite-energy modes, with the threshold of
Eq.~(\ref{eq:threshold}), is a result of the model, whereas that this particular
pair of channels realises it is a proposal, the numbers of which follow from a
prescribed modulation rather than from the coupled two-channel electrostatics.
The selection of the pumping wavenumber, which might have been thought the weak
link, is secured by the same viscosity that limits the crystal, the overtones of
the cavity comb being damped quadratically while the gain is not. What is not
secured is the length: a cavity short enough to oscillate is two orders of
magnitude shorter than the crystal that radiates, Eq.~(\ref{eq:Lmax}), so that
the pump must be an array of some hundreds of phase-locked cavities, and whether
such an array can be locked to the tolerance that the zone-centre condition
imposes is the question upon which the whole architecture now turns.

\begin{acknowledgments}
It is a pleasure to dedicate this work to J. T. Mendon\c ca, whose papers on time refraction have greatly inspired this work.
\end{acknowledgments}

\bibliographystyle{apsrev4-2}
\bibliography{REFS}
\end{document}